# Study of pnictides for photovoltaic applications

Jayant Kumar and Gopalakrishnan Sai Gautam*

Department of Materials Engineering, Indian Institute of Science, Bengaluru 560012, India

*E-mail: saigautamg@iisc.ac.in

**Abstract**

For the transition into a sustainable mode of energy usage, it is important to develop photovoltaic materials that exhibit better solar-to-electricity conversion efficiencies, a direct optimal band gap, and made of non-toxic, earth abundant elements compared to the state-of-the-art silicon photovoltaics. Here, we explore the non-redox-active pnictide chemical space, including binary $A_3B_2$, ternary $AA'_2B_2$, and quaternary $AA'A''B_2$ compounds (A, A', A'' = Ca, Sr, or Zn; B = N or P), as candidate beyond-Si photovoltaics using density functional theory calculations. Specifically, we evaluate the ground state configurations, band gaps, and 0 K thermodynamic stability for all 20 pnictide compositions considered, besides computing the formation energy of cation vacancies, anion vacancies, and cation anti-sites in a subset of candidate compounds. Importantly, we identify $SrZn_2N_2$, $SrZn_2P_2$, and $CaZn_2P_2$ to be promising candidates, exhibiting optimal (1.1-1.5 eV) hybrid-functional-calculated band gaps, stability at 0 K, and high resistance to point defects (formation energies >1 eV), while other possible candidates include $ZnCa_2N_2$ and $ZnSr_2N_2$, which may be susceptible to N-vacancy formation. We hope that our study will contribute to the practical development of pnictide semiconductors as beyond-silicon light absorbers.

# 1 Introduction

The 21$^{st}$ century world is in need of a dramatic shift in the energy sector, given the increasingly unsustainable nature of fossil fuel usage and the associated climate change. Among renewable sources, solar energy, i.e., the conversion of solar radiation into electricity via photovoltaics (PVs), has significant potential in reducing our fossil fuel usage.[1] Notably, the Shockley-Queisser limit for maximum efficiency in single-junction PV devices requires semiconductors with a direct band gap in the 1.1-1.5 eV range.[2] Commercially, PVs are typically made based on crystalline Si, and do require larger material quantities (i.e., PV panels are thick) and higher manufacturing hosts, necessitated by the indirect band gap (1.12 eV) of Si.[3] Thus, engineering and discovering direct gap semiconductors, which exhibit a 1.1-1.5 eV band gap, are made out of environmentally sustainable and non-toxic elements, and are reasonably cheap to manufacture, is still an active area of research.[4,5,6]

Semiconductors that have been explored as potential beyond-Si photovoltaics, which are often compounds, include (In,Ga)As, CdTe, $Cu_2ZnSnS_4$ (CZTS), $Cu_2InGaSe_4$ (CIGS), and their doped counterparts. There are several challenges in utilizing the aforementioned compounds as PVs, including some of the elements being toxic (e.g., As in (In,Ga)As), lack of abundance of elements (e.g., Te in CdTe[7]), difficulties in synthesis of phase-pure compounds resulting in the presence of secondary phases (e.g., secondary phases of Cu-Se and Ag-Cu-Se forms in (Ag,Cu),(In,Ga)Se$_2$ [8]) and the spontaneous formation of detrimental point defects (e.g., $Cu_{Zn} + Zn_{Cu}$ and $Sn_{Zn} + 2Cu_{Zn}$ anti-site clusters in CZTS). In particular, point defects that form either during synthesis or processing (e.g., high temperature annealing) can be harmful in reducing PV performance, by introducing structural distortions, altering the band gap, and/or formation of deep trap states within the band gap. Hence, it is important to evaluate the tendency to form intrinsic point defects while considering any novel semiconductor compound as a PV candidate.

Here, we explore a set of 20 pnictide compounds, including binary, ternary, and quaternary nitrides and phosphides, as potential PV candidates, using density functional theory (DFT[9,10]) calculations. Specifically, we explore pnictides of the formula, $A_3B_2$, $AA'_2B_2$, and $AA'A''B_2$ where A, A' and A'' are permutations of divalent cations $Ca^{2+}$, $Sr^{2+}$, and $Zn^{2+}$ and B is either $N^{3-}$ or $P^{3-}$. Our choice of divalent cations is largely motivated by a few existing studies that have explored pnictides containing these elements as possible PVs, such as $Ca_3N_2$, $Zn_3N_2$, $Zn_3P_2$, and $CaZn_2N_2$,[11,12,13,14,15] while a widespread theoretical or experimental screening for

PV candidates has not been done so far in the pnictide chemical space.[16] Moreover, Ca, Sr, and Zn are reasonably abundant on the earth's crust and non-toxic.

Apart from determining the ground state atomic configuration in disordered compounds, we evaluate their suitability as a PV material by calculating their band gap, 0 K thermodynamic stability, and the formation energies of select point defects. Importantly, we identify five potential candidates, namely, $ZnCa_2N_2$, $SrZn_2N_2$, $ZnSr_2N_2$, $CaZn_2P_2$, and $SrZn_2P_2$, which exhibit an optimal (1.1-1.5 eV) band gap and are thermodynamically stable. Among the candidates that we have identified, $CaZn_2N_2$ and $SrZn_2N_2$ have been experimentally reported to exhibit band gaps of 1.90 eV and 1.60 eV, respectively.[14,17] Notably, considering point defect formation energies, we find $SrZn_2N_2$, $CaZn_2P_2$, and $SrZn_2P_2$ to be particularly promising since they exhibit a formation energy of >1 eV for cation vacancies, anion vacancies, and cation anti-sites. We hope that our study will reinvigorate research in the pnictide chemical space for potential PV materials, and will aid in the development of high efficiency, beyond-Si PVs that consist of sustainable and non-toxic constituents.

## 2 Methods

We calculated the total energies and electronic structures, for all pnictides considered, using DFT as implemented in the Vienna ab initio simulation package[18,19] and employing the projector-augmented-wave theory[20] (list of potentials used is provided in the Supporting Information—SI). We used an energy cut-off of 520 eV on a plane-wave basis for all calculations and sampled the irreducible Brillouin zone using Γ-centred Monkhorst-Pack[21] meshes with a density of 32 $k$-points per Å. We relaxed all structures by allowing the cell shape, cell volume, and ionic positions to change, without preserving symmetry, until the atomic forces and total energies reduced below |0.01| eV/Å, and $10^{-5}$ eV, respectively. We treated the electronic exchange-correlation interactions with the strongly constrained and appropriately normed (SCAN[22]) functional for all structure relaxation calculations. For the electronic density of states (DOS) calculations, we considered the converged ground state structure for each composition, and used the "fake" self-consistent field (SCF) procedure, with a mesh density of 64 $k$-points per Å. Note that the set of $k$-points sampled during the structure relaxation were retained with their original weights, while the newly introduced $k$-points were sampled with zero weights within the fake-SCF procedure. Since SCAN typically underestimates the band gap of semiconducting systems[3,23–27], we also performed DOS calculations using the Heyd-

Scuseria-Ernzerhof (HSE06[28]) hybrid functional, using an identical fake-SCF procedure and the SCAN-relaxed structures.

The initial crystal structure for a subset of the compounds considered in this work (see **Table 1**), were obtained from the inorganic crystal structure database (ICSD).[29] For the remaining compounds, we derived the initial configurations based on the available unique structures of other compounds (i.e., unique space groups) via ionic substitution (see **Section 3.1** for details). For structures that exhibited disorder within the cation and/or anion sub-lattices, we enumerated symmetrically distinct configurations using the OrderDisorderedStructureTransformation() class within the pymatgen package.[30] Note that the electronic structure, thermodynamic stability, and defect calculations were performed only for the ground state configuration (as evaluated by DFT) for each compound. To evaluate the thermodynamic stability, we calculated the 0 K convex hulls of the quaternary Ca-Sr-Zn-N and Ca-Sr-Zn-P systems, which includes the elements, and all possible binary, ternary, and quaternary ordered structures, as available in the ICSD.

For the five candidates, namely $ZnCa_2N_2$, $ZnSr_2N_2$, $SrZn_2N_2$, $CaZn_2P_2$, and $SrZn_2P_2$, we calculated the defect formation energies of cation vacancies, anion vacancies, and cation anti-sites. We used 3×3×2 (90-atom) supercells of $CaZn_2P_2$, $SrZn_2P_2$, and $SrZn_2N_2$ and 3×3×2 (180-atom) supercells of $ZnCa_2N_2$[14] and $ZnSr_2N_2$ to model the defective structures. The formation energy of any defect ($E^f$) is given by:[31]

$$E^f = E_{defect} - E_{bulk} - \Sigma_i n_i \mu_i + qE_F + E_{corr}$$

where $E_{defect}$ and $E_{bulk}$ are the total energies of supercell with and without defect, respectively, $n_i$ represents the number of $i$-species atoms added (> 0) and/or removed (< 0) to form the defect with $\mu_i$ the corresponding chemical potential. We obtained the range of relevant $\mu_i$ for all vacancy defects considered from our 0 K convex hulls (values listed in **Table S1** of SI). The terms $qE_F$ and $E_{corr}$ indicate the exchange of charge(s) with the Fermi energy ($E_F$) of the pristine semiconductor and the corresponding correction term, which are relevant for the formation of charged defects. However, we have only considered neutral defects ($q = E_{corr} = 0$), which corresponds to removing (or adding) the entire atom to form a defect.

# 3 Results

## 3.1 Structures

**Table 1**: Details of structures obtained from ICSD, with their corresponding space groups. O and D indicate ordered and disordered structures, respectively. The number of unique orderings obtained upon enumeration are listed for disordered structures. The Wyckoff positions of unique cation sites are displayed in the final column on the right.

| Compounds | Space group | Ordered/Disordered (from ICSD) | Cations/total atoms | Number of unique orderings | Occupancy (cations per site) | Unique cation sites |
|---|---|---|---|---|---|---|
| $Ca_3N_2$ | $Ia\bar{3}$ | O | 24/40 | 1 | Ca: 1.0 | 48e |
| $Zn_3N_2$ | | | | | Zn: 1.0 | |
| $Ca_3P_2$ | $P6_3/mcm$ | D | 9/15 | 2 | Ca: 0.9 | 4d, 6g |
| $Sr_3P_2$ | $I\bar{4}3d$ | | 24/40 | 18 | Sr: 1.0 | 16c |
| $Zn_3P_2$ | $P4_2/nmc$ | | | | Zn: 1.0 | 8g, 8g, 8g |
| $CaZn_2P_2$ | $P\bar{3}m1$ | O | 3/5 | 1 | Ca: 1.0, Zn: 1.0 | 1a, 2d |
| $SrZn_2P_2$ | | | | | Sr: 1.0, Zn: 1.0 | |
| $ZnSr_2N_2$ | $I4/mmm$ | | 6/10 | | Zn: 1.0, Ca: 1.0 | 2a, 4e |
| $ZnCa_2N_2$ | | | | | | |
| $ZnSr_2P_2$ | $P6_3/mmc$ | D | 3/5 | | Zn: 0.5, Sr: 1.0 | 2a, 2d |

Among the 20 pnictides considered in this work, the ICSD contains structures of 10 compounds, with the structural data compiled in **Table 1** and the initial structures displayed in **Figure 1**. We find that $Ca_3N_2$ and $Zn_3N_2$ have an anti-bixbyite structure (space group: $Ia\bar{3}$) derived from bixbyite-$Mn_2O_3$ (**Figure 1a**).[32] $Zn_3P_2$ (**Figure 1e**) and $ZnSr_2N_2$ (**Figure 1f**) crystallize in tetragonal lattices, but with different space groups, namely $P4_2/nmc$ and $I4/mmm$, respectively, with the structures of $ZnSr_2N_2$ and $ZnCa_2N_2$ being identical (i.e., Ca and Sr occupy identical cation sites in a tetragonal lattice).[14,33,34] $CaZn_2P_2$ and $SrZn_2P_2$ exhibit hexagonal ($P\bar{3}m1$) structures (**Figure 1d**), while $Ca_3P_2$ (**Figure 1b**) and $ZnSr_2P_2$ (**Figure 1g**) are also hexagonal with a different space group ($P6_3/mmc$).[35–37] Among binary compounds, $Ca_3P_2$ and $Sr_3P_2$ (**Figure 1c**) exhibit disorder in the cation and anion lattices, respectively, where we have used the 3×1×1 (40-atom) supercell of the $I\bar{4}3d$-primitive cell of $Sr_3P_2$ to enumerate the P orderings. In ternaries, only $ZnSr_2P_2$ exhibits disorder within the Zn-sites. Upon enumeration, we obtain 2, 18, and 1 symmetrically distinct configurations in $Ca_3P_2$, $Sr_3P_2$, and $ZnSr_2P_2$, respectively, with the initial structures of the ground state configurations displayed in **Figure 1**.

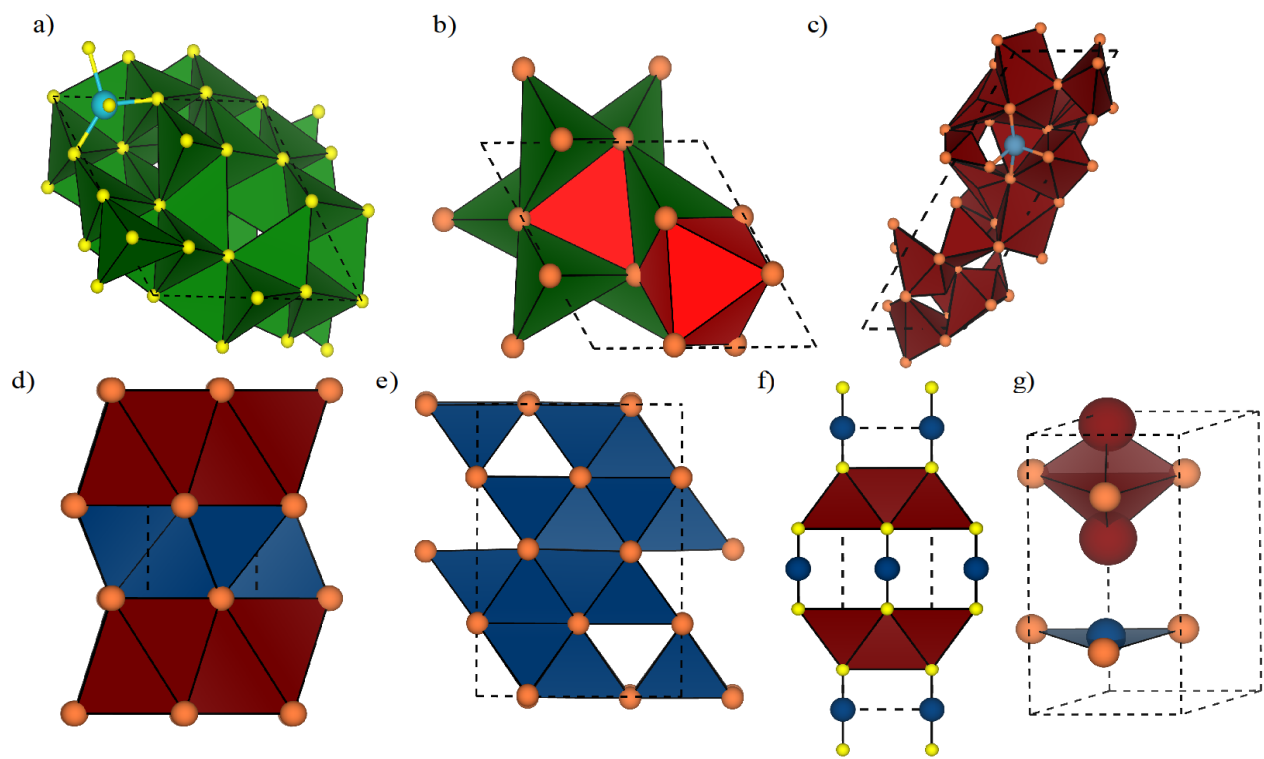

**Figure 1**: Initial structures of nitrides and phosphides, as available in the ICSD, including a) $Ca_3N_2$ (and $Zn_3N_2$), b) $Ca_3P_2$, c) $Sr_3P_2$, d) $SrZn_2P_2$ (and $CaZn_2P_2$), e) $Zn_3P_2$, f) $ZnSr_2N_2$ (and $ZnCa_2N_2$), and g) $ZnSr_2P_2$. The open polyhedron (panel a) shows the tetrahedral coordination of the Ca (or Sr) atom in $Ca_3N_2$ (in $Sr_3P_2$). Different shades of polyhedra in $Ca_3P_2$ (panel b) show the different coordination environments of tetrahedra (dark green) and octahedra (red) for Ca atom. Color codes: Ca – green, Sr – brown, Zn – blue, N – yellow, and P – orange. In panels d and f, Ca and Sr occupy identical sites in the corresponding Sr- and Ca-containing compounds.

For compounds whose structures are not available in the ICSD, we used the ordered structures with unique space groups in **Table 1** as templates to derive the new initial structures. Specifically, we used the unit cells of $Zn_3P_2$ ($P4_2/nmc$), $ZnCa_2N_2$/$ZnSr_2N_2$ ($I4/mmm$), and $CaZn_2P_2$/$SrZn_2P_2$ ($P\bar{3}m1$) as the templates, all of which exhibit 3 distinct cation sites that can be interchanged. For example, substituting all Ca with Sr in $CaZn_2P_2$ (1a sites) and $ZnCa_2N_2$ (4e sites) lead to identical Sr-containing structures, namely, $SrZn_2P_2$ and $ZnSr_2N_2$, respectively. Thus, Ca and Sr sites can be used for the occupation of either cation. Römer et. al. studied different theoretical structures for $Sr_3N_2$, such as anti-B-sesquioxide, anti-$Rh_2O_3$-II, hexagonal $P6_3/mmc$, and anti-bixbyite $Ia\bar{3}$.[38] Among these structures, the anti-bixbyite was found to have the lowest energy, similar to the binary Ca and Zn nitrides. Hence, we used the $Ca_3N_2$ structure as a template to initialize the anti-bixbyite $Sr_3N_2$ structure by substituting Ca with Sr.

For generating the ternary nitride structures, we did not use the $Ia\bar{3}$-$Ca_3N_2$ or $Zn_3N_2$ structures as templates since the framework contains 24 distinct cation sites, which presents a combinatorial limitation (i.e., large number of unique configurations to consider) to computational substitution. Similarly, we did not consider the $Sr_3P_2$, $Ca_3P_2$, and $ZnSr_2P_2$ disordered structures as templates due to the combinatorial limitation. The DFT-calculated energies for the different initial configurations considered and the initial configurations of the ground states for all theory-derived structures are compiled in **Table S2** and **Figure S1** of the SI.

As an example of our procedure for theoretically deriving structures, consider the case of $CaSr_2P_2$. Here, we generated three sets of structures, by substituting Zn and N with Ca and P in $ZnSr_2N_2$ (i.e., the Sr sites left as-is), replacing Sr and Zn with Ca and Sr in $SrZn_2P_2$ (i.e., 1a sites are occupied by Ca and 2d sites are occupied by Sr after replacement, which ensures correct stoichiometry of Ca and Sr in $CaSr_2P_2$), and exchange two 8g sites of Zn with Sr and one 8g site with Ca in $Zn_3P_2$. The aforementioned substitutions gave rise to 1, 1, and 3 symmetrically distinct configurations of $CaSr_2P_2$ as derived from $ZnSr_2N_2$, $SrZn_2P_2$, and $Zn_3P_2$, respectively, out of which the DFT-calculated ground state was one of the configurations derived from $Zn_3P_2$ (**Figure S1**). Among the theoretical ternaries (see **Table S2** and **Figure S1**), the DFT-calculated ground state configurations of $ZnCa_2P_2$ and $CaSr_2N_2$ were derived from $ZnCa_2N_2$ ($I4/mmm$), and $Zn_3P_2$ ($P4_2/nmc$), respectively, while the ground states of the other ternaries ($CaZn_2N_2$, $SrZn_2N_2$, $SrCa_2N_2$, and $SrCa_2P_2$) were based on $CaZn_2P_2/SrZn_2P_2$ ($P\bar{3}m1$). Notably, our ground state configuration of $SrZn_2N_2$ and $CaZn_2N_2$ (derived from a $P\bar{3}m1$ space group) are consistent with previous theoretical studies as well,[17] highlighting the robustness of our substitution strategy.

For the quaternaries $CaSrZnN_2$ and $CaSrZnP_2$ (see **Table S3 a**nd **Figure S1**), we derived the theoretical structures by substitution that was done analogously to the ternary compounds. For example, to obtain $CaSrZnP_2$ configurations, we substituted 1 8g site each with Sr and Ca in $Zn_3P_2$, considered Ca, Sr, and Zn occupying the 1a and 2d sites in $CaZn_2P_2$ (i.e., each cation is allowed to occupy one of the distinct Wyckoff sites), and the three cations residing in the 2a and 4e sites and all N replaced with P in $ZnSr_2N_2$. Such substitution resulted in 6, 3, and 3 unique configurations of $CaSrZnP_2$ derived from $Zn_3P_2$, $CaZn_2P_2$, and $ZnSr_2N_2$, respectively, out of which the DFT-calculated ground state was based on the ternary-$CaZn_2P_2$. The $ZnSr_2N_2$ ternary formed the ground state template for $CaSrZnN_2$ quaternary.

## 3.2 Band gaps

**Table 2**: Band gaps ($E_g$ in eV) of the pnictides considered in this work, as calculated using the SCAN and HSE06 functionals. Expt. indicates experimental data. The structural origin column indicates whether the initial structure was obtained from ICSD or via theoretical substitution of atoms.

| Compounds | Structural origin | $E_g$ (eV) | | |
|---|---|---|---|---|
| | | SCAN | HSE06 | Expt. |
| Binaries | | | | |
| $Ca_3N_2$ | ICSD | 1.19 | 1.75 | 1.90 [11] |
| $Ca_3P_2$ | ICSD | 0.35 | 0.59 | - |
| $Sr_3N_2$ | Theoretical | 0.40 | 0.93 | - |
| $Sr_3P_2$ | ICSD | 0.12 | 0.38 | - |
| $Zn_3N_2$ | ICSD | Metallic | 0.64 | 1.23 [39] |
| $Zn_3P_2$ | ICSD | 0.38 | 0.90 | 1.46 [13] |
| Ternaries | | | | |
| $CaSr_2N_2$ | Theoretical | 0.71 | 1.22 | - |
| $CaSr_2P_2$ | Theoretical | 1.46 | 1.75 | - |
| $SrZn_2N_2$ | Theoretical | 0.67 | 1.39 | 1.60 [17] |
| $SrZn_2P_2$ | ICSD | 0.84 | 1.18 | - |
| $CaZn_2N_2$ | Theoretical | 0.88 | 1.65 | 1.90 [14] |
| $CaZn_2P_2$ | ICSD | 0.88 | 1.23 | - |
| $SrCa_2N_2$ | Theoretical | 1.85 | 2.36 | - |
| $SrCa_2P_2$ | Theoretical | 1.73 | 2.12 | - |
| $ZnCa_2N_2$ | ICSD | 0.74 | 1.40 | 1.60 [14] |
| $ZnCa_2P_2$ | Theoretical | 0.36 | 0.80 | - |
| $ZnSr_2N_2$ | ICSD | 0.89 | 1.38 | - |
| $ZnSr_2P_2$ | ICSD | Metallic | 0.12 | - |
| Quaternaries | | | | |
| $CaSrZnN_2$ | Theoretical | 0.77 | 1.33 | - |
| $CaSrZnP_2$ | Theoretical | 1.01 | 1.41 | - |

All band gap ($E_g$) data, including the HSE06-calculated, SCAN-calculated, and available experimental values, are compiled in **Table 2**, which also lists whether the calculated DOS is originally from ICSD or a theoretical structure (see **Section 3.1**). In the case of $SrZn_2N_2$ and $CaZn_2N_2$, the $E_g$ has been measured experimentally[14,17], and the structures reported to be trigonal ($P\bar{3}m1$), in accordance with our results. However, the reported structures of $SrZn_2N_2$ and $CaZn_2N_2$ are not available in the ICSD, due to which we utilized our template+substitution strategy to obtain initial structures for our calculations. Notably, we find that all our SCAN-calculated $E_g$ (in **Table 2**) are lower than HSE06-calculated $E_g$, which is expected since the inclusion of exact exchange in hybrid functionals facilitates electron localization and hence result in larger $E_g$ than semi-local functionals (such as SCAN). There are also cases, namely $Zn_3N_2$ and $ZnSr_2P_2$, where SCAN predicts a qualitatively different (i.e., metallic) electronic structure compared to HSE06, but such compounds may not be candidates for solar absorbers since their actual $E_g$ may be low (i.e., < 1 eV).

Interestingly, there are cases of nitrides exhibiting larger $E_g$ than phosphides of the same cation composition (e.g., $Ca_3N_2$ vs. $Ca_3P_2$, and $ZnCa_2N_2$ vs. $ZnCa_2P_2$), while the vice-versa is also true in some cases (e.g., $Zn_3P_2$ vs. $Zn_3N_2$, and $CaSr_2P_2$ vs. $CaSr_2N_2$). Importantly, the HSE06-$E_g$ are in better agreement (but underestimating) with respect to the available experimental $E_g$ than SCAN. The level of $E_g$ underestimation by HSE06 is severe in binary pnictides (48% and 38% in $Zn_3N_2$ and $Zn_3P_2$, respectively), compared to the ternaries (13%, 13%, and 12.5% in $SrZn_2N_2$, $CaZn_2N_2$, and $ZnCa_2N_2$, respectively). Thus, for the selection of candidates with suitable band gaps for PV applications, we can still use the general range of 1.1-1.5 eV to ensure that there are no false negatives among the compositions considered.

Applying the 1.1-1.5 eV range within our HSE06-$E_g$, we find 6 possible ternary pnictides as candidate PVs, namely $CaSr_2N_2$ ($E_g$ ~1.22 eV), $SrZn_2N_2$ (1.39 eV), $SrZn_2P_2$ (1.18 eV), $CaZn_2P_2$ (1.23 eV), $ZnCa_2N_2$ (1.40 eV), and $ZnSr_2N_2$ (1.38 eV), apart from the quaternaries, $CaSrZnN_2$ (1.33 eV) and $CaSrZnP_2$ (1.41 eV). Our calculated band gap in $ZnSr_2N_2$ (1.38 eV) is in agreement with a previous study utilizing the modified Becke-Johnson potential[40] (~1.45 eV[41]), although our predicted value for $ZnCa_2N_2$ (1.40 eV) is quite different from the same report (~1.72 eV[41]). **Figure 2** compiles the HSE06-calculated DOS for the six candidate ternaries, while the remaining HSE06 and SCAN-calculated DOS for all pnictides considered are displayed in the SI (**Figures S2-S6**). In all our DOS plots, the orange lines correspond to the anion (i.e., N or P) $p$ states, while the red, green, purple, and/or brown lines correspond to the cation $s$ (for Ca, Sr, and Zn) and/or $d$ (for Zn only) states. The dotted blue lines in our DOS panels indicate the valence and conduction band edges, while the Fermi level in metallic systems is displayed by dashed black lines. The zero on the energy scale in all DOS visualizations is set to the valence band maximum (VBM) in gapped systems, while the zero is set to the Fermi level in metallic systems.

Among the candidate ternaries, the VBM consists almost exclusively of the anionic $p$ states, with low overlap of cationic $s$ or $d$ states, suggesting that the cation-anion bonds are quite ionic in nature. While anionic $p$ states dominate the conduction band minimum (CBM) in $CaSr_2N_2$ (**Figure 2a**), $ZnCa_2N_2$ (**Figure 2e**) and $ZnSr_2N_2$ (**Figure 2f**), the Zn $s$ states do contribute significantly to the CBM in $SrZn_2N_2$ (**Figure 2b**), $SrZn_2P_2$ (**Figure 2c**), and $CaZn_2P_2$ (**Figure 2d**). In the case of quaternaries, anionic $p$ states dominates the VBM in both compounds, while N $p$ and Zn $s$ states dominate CBM in $CaSrZnN_2$ (**Figure S5**) and $CaSrZnP_2$ (**Figure S6**) respectively. Among the ternary candidates identified in our work, $SrZn_2N_2$,

ZnSr$_2$N$_2$, and ZnCa$_2$N$_2$ have been explored as solar absorbers before (i.e., $E_g$ calculated and/or measured),[14,17] while the remaining compounds (CaSr$_2$N$_2$, SrZn$_2$P$_2$, and CaZn$_2$P$_2$) have not been studied as PVs so far. However, whether these compounds are feasible PV candidates depends on their thermodynamic stability (**Section 3.3**) and intrinsic tendency to form detrimental point defects (**Section 3.4**). Note that although the experimental band gaps of Zn$_3$N$_2$ and Zn$_3$P$_2$ are in the optimal range (1.23 eV and 1.46 eV, respectively), previous attempts at utilizing Zn$_3$P$_2$ as solar absorbers have yielded poor efficiencies, namely ~4% with expected Mg-doped improvements reaching 8-10% in Zn$_3$P$_2$.[42] Low efficiency in Zn$_3$P$_2$ have been attributed to difficulties in synthesizing *p-n* homojunctions of Zn$_3$P$_2$[42], while the high moisture sensitivity and possible oxygen contamination of Zn$_3$N$_2$ has led to large variations in measured band gaps (1.06-3.2 eV),[43] with possible improvements in band gap engineering arising out of synthesizing Zn$_{3-3x}$Mg$_{3x}$N$_2$ alloys.[44] Hence, we will not be considering Zn$_3$N$_2$ and Zn$_3$P$_2$ as candidate materials in the rest of the manuscript.

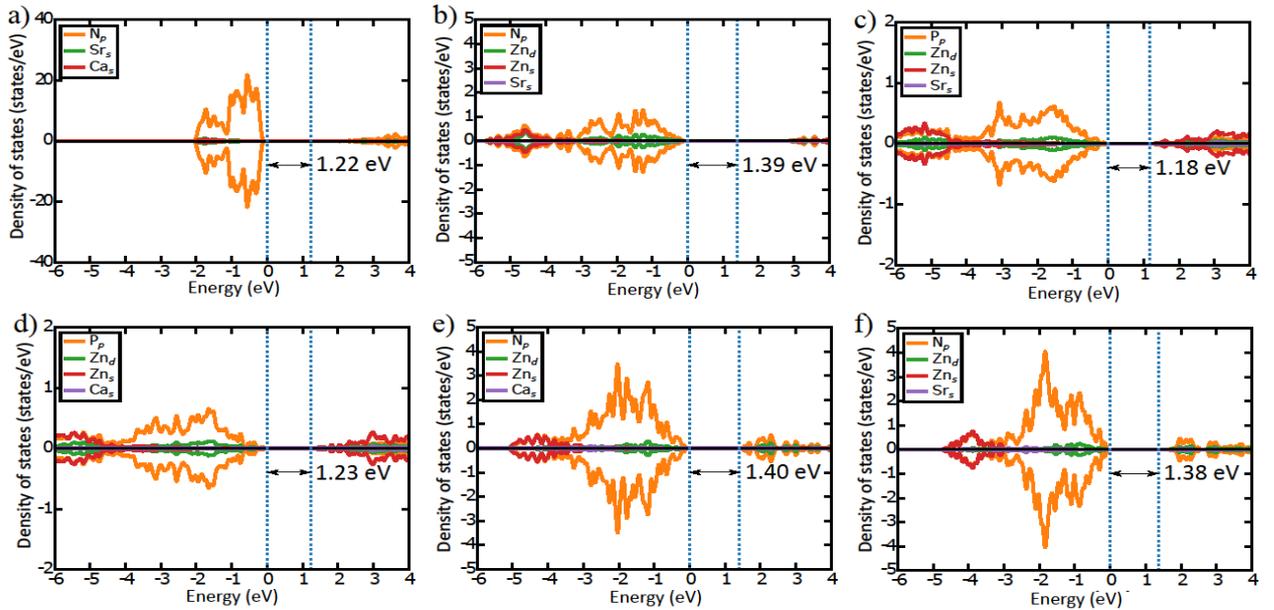

**Figure 2**: HSE06 calculated DOS for ternary candidate pnictides, including (a) CaSr$_2$N$_2$, (b) SrZn$_2$N$_2$, (c) SrZn$_2$P$_2$, (d) CaZn$_2$P$_2$, (e) ZnCa$_2$N$_2$, and (f) ZnSr$_2$N$_2$. Text annotation in each panel indicates the calculated band gap with the dotted blue lines highlighting the band edges.

## 3.3 Thermodynamic stability

We calculated the 0 K convex hulls for the quaternary systems, namely Ca-Sr-Zn-N and Ca-Sr-Zn-P, comprising of all the ordered structures available on ICSD (see list compiled in **Table S3** and the theoretical/enumerated ground states that we have considered. For ease of visualization, we have compiled three ternary projections for each quaternary system, namely, Ca-Sr-N, Ca-Zn-N, and Sr-Zn-N (panels **a-c**) for Ca-Sr-Zn-N quaternary, and Ca-Sr-P, Ca-Zn-P, and Sr-Zn-P (panels **d-f**) for Ca-Sr-Zn-P quaternary, in **Figure 3**. All compounds that are stable at 0 K are indicated by filled circles in Figure 3, while the hollow circles indicate unstable entities. The stable (unstable) pnictides considered in this work are highlighted by green (red) filled (hollow) circles and corresponding text annotations in **Figure 3**. Note that previous theoretical studies have utilized a ~30 meV/atom energy above the hull ($E^{hull}$) threshold value,[45] beyond which synthesis of metastable/unstable compounds may become difficult experimentally.

We find all the ICSD-derived structures to be stable at 0 K, which is in line with expectations that the ICSD-derived structures have been experimentally characterized, with the only exception being $ZnSr_2P_2$ that is ~90 meV/atom unstable compared to other phases ($Sr_3P_2$ and $SrZn_2P_2$) in the Zn-Sr-P phase diagram (**Figure 3f**). Note that the experimental structure of $ZnSr_2P_2$ is disordered, which typically results in an excess of configurational entropy that may stabilize the compound at higher temperatures, thus facilitating its synthesis. Among the theoretical structures, the stable pnictides are $SrZn_2N_2$, $CaZn_2N_2$, and $SrCa_2P_2$, out of which $SrZn_2N_2$ and $CaZn_2N_2$ have been reported to be synthesized before.[14,17] Thus, we find $SrCa_2P_2$ to be a stable compound that has not been experimentally reported nor theoretically studied in detail so far.

The unstable ternaries among the pnictides considered are, $CaSr_2N_2$ ($E^{hull}$~72 meV/atom), $CaSr_2P_2$ (65 meV/atom), $SrCa_2N_2$ (34 meV/atom), and $ZnCa_2P_2$ (45 meV/atom) with the theoretical binary, $Sr_3N_2$ also marginally unstable ($E^{hull}$~9 meV/atom). Both the theoretical quaternaries included in this work, $CaSrZnN_2$ and $CaSrZnP_2$ are also unstable ($E^{hull}$ ~34 meV/atom and 68 meV/atom, respectively). Given the band gap data of **Table 1** and **Figure 2**, and including the 0 K thermodynamic stability as an additional filter, the set of viable PV candidates reduces to 5 compounds, $SrZn_2N_2$, $SrZn_2P_2$, $CaZn_2P_2$, $ZnCa_2N_2$, and $ZnSr_2N_2$, thus eliminating $CaSr_2N_2$ as a candidate due to its thermodynamic instability. Note that although metastable compounds (e.g., $CaSrZnN_2$) may be synthesizeable, such compounds usually exhibit a higher tendency to form point defects than thermodynamically stable compounds.[46] Hence, for the calculation of point defect energies, we restrict ourselves to the above 5 candidates.

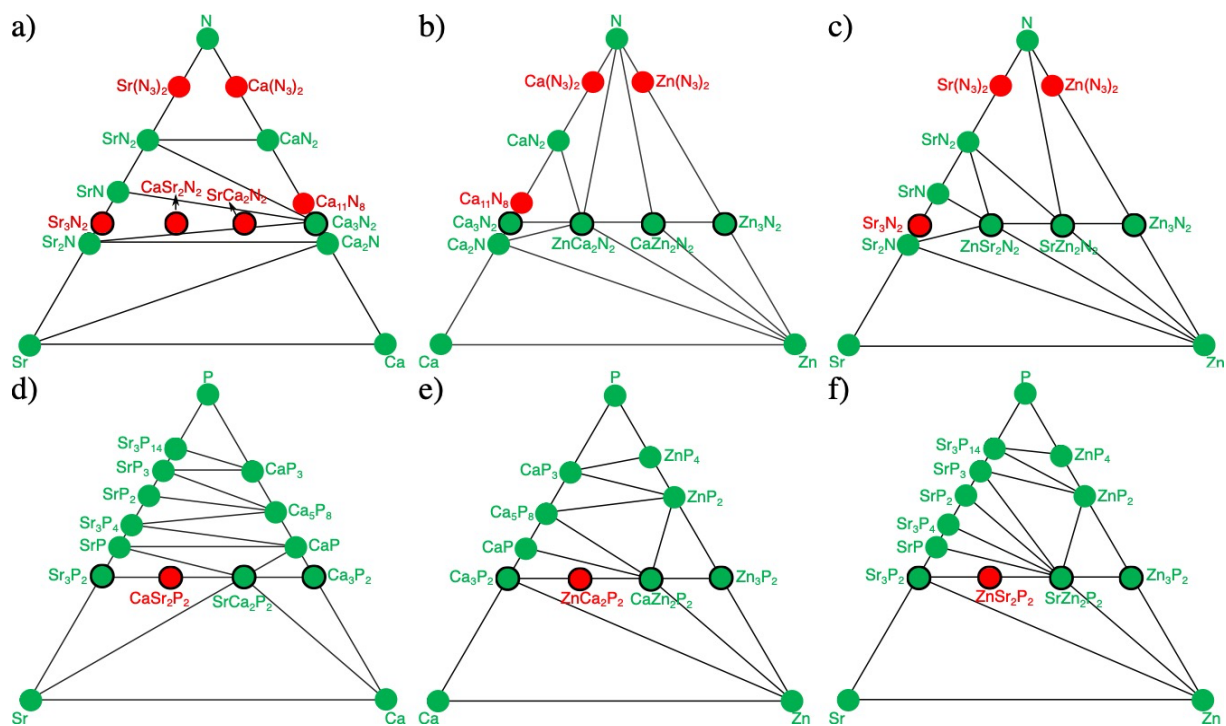

**Figure 3**: SCAN-calculated 0 K ternary phase diagrams, including (a) Ca-Sr-N, (b) Ca-Zn-N, (c) Sr-Zn-N, (d) Ca-Sr-P, (e) Ca-Zn-P, and (f) Sr-Zn-P systems. Green and red circles (and corresponding text annotations) are stable and unstable/metastable compounds, respectively. Black outlined circles represent candidate compounds considered in this work.

### 3.4 Point defect formation energies

For the five candidates identified from our band gap and stability calculations, namely, $SrZn_2N_2$, $SrZn_2P_2$, $CaZn_2P_2$, $ZnCa_2N_2$, and $ZnSr_2N_2$, we evaluate the point defect energetics, as displayed in **Figure 4**. Specifically, we consider the formation of A and A' cation vacancies (i.e., $Vac_A$ and $Vac_{A'}$ where Vac is vacancy, yellow and green bars in **Figure 4**), anion vacancies ($Vac_B$, orange bar), and cation anti-site clusters ($A_{A'}+A'_A$, blue bar). For example, in the case of $SrZn_2N_2$, we calculate the energy of formation for $Vac_{Sr}$, $Vac_{Zn}$, $Vac_N$, and $Sr_{Zn}+Zn_{Sr}$ (i.e., Sr occupying a Zn site+Zn occupying a Sr site). Hashed regions in **Figure 4** indicate the variation in the cation/anion vacancy formation energies with the corresponding change in the $\mu$ of the species being removed. The range of $\mu$ considered is based on the 0 K phase diagrams, i.e., within the region of thermodynamic stability of the pnictide considered. In the case of anti-sites, we created a cluster by exchanging nearest possible neighbors of A and A' atoms. The dashed black line in **Figure 4** indicates a 1 eV threshold of point defect formation energies. Although it is preferable for point defect formation energies to be as high as possible in pristine bulk phases, we use 1 eV as an arbitrary threshold to account for high-temperature synthesis protocols that may be used during fabrication. Note that this 1 eV threshold can be changed, if necessary.

Importantly, we find that all the five candidates exhibit anti-site cluster and cation vacancy formation energies well above the 1 eV threshold (**Figure 4**), ranging from ~1.42 eV for $Vac_{Ca}$ in $CaZn_2P_2$ to 5.25 eV for $Vac_{Ca}$ in $ZnCa_2N_2$, highlighting the large degree of resistance that the candidate pnictides exhibit for formation of such defects. In terms of anion vacancies, while $SrZn_2N_2$ (2.29 – 2.64 eV), $SrZn_2P_2$ (2.02 – 3.26 eV), and $CaZn_2P_2$ (1.88 - 3.17 eV) exhibit formation energies above the 1 eV threshold, $ZnCa_2N_2$ (0.42 - 2.14 eV) and $ZnSr_2N_2$ (0.47 – 2.00 eV) do not, indicating that the former three compounds exhibit better resistance to *n*-type anion vacancies, especially under anion-poor conditions. Thus, combining the band gap, thermodynamic stability, and high degree of resistance to point defect energetics, we find $SrZn_2N_2$, $SrZn_2P_2$, and $CaZn_2P_2$ to be the most promising candidate materials for PV applications. In case low temperature synthesis and/or heat treatment protocols are utilized, we believe that $ZnCa_2N_2$ and $ZnSr_2N_2$ may also be relevant for PVs, since their defect formation energies are higher than the values reported for common point defects in kesterite-based PV materials.[3,26,47,48] While $SrZn_2N_2$ has been explored for its semiconducting properties before[17], $SrZn_2P_2$ and $CaZn_2P_2$ are novel candidates whose electronic properties have not been explored *a priori*.

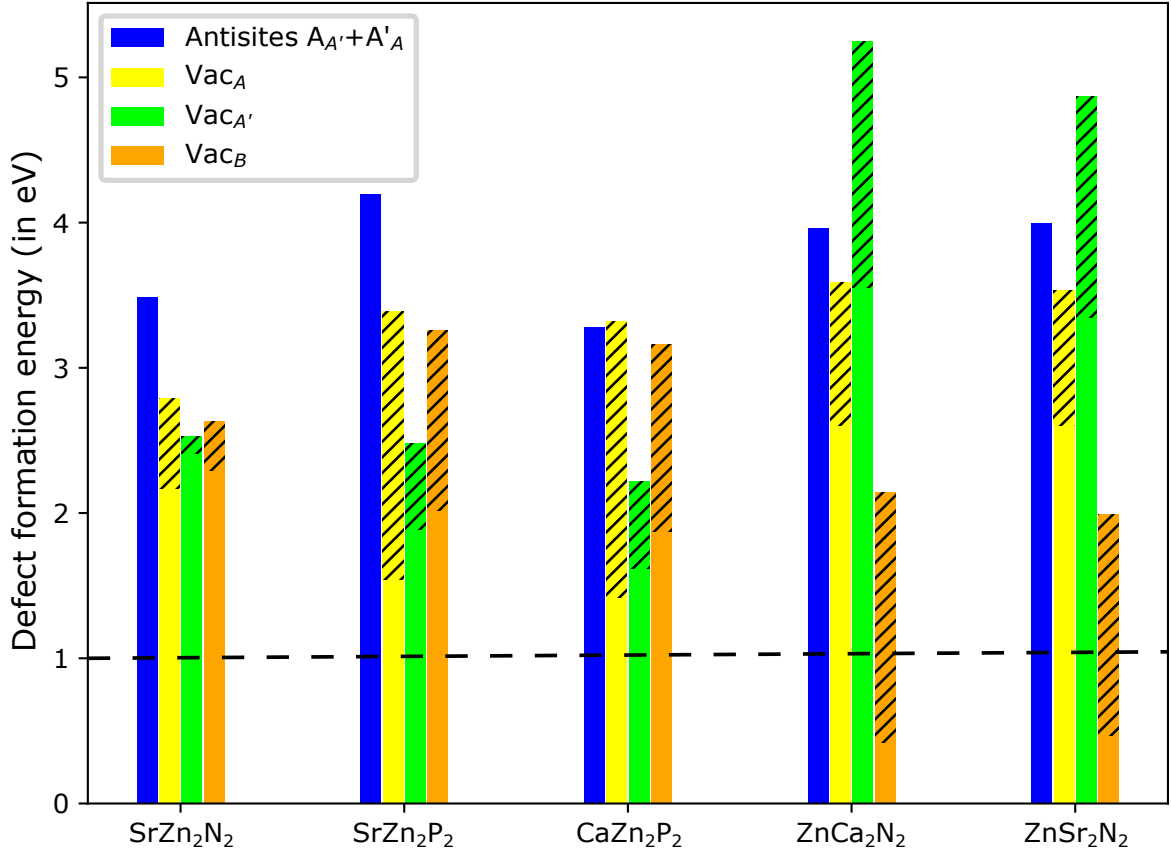

**Figure 4**: Point defect formation energies for A-site and A'-site cation vacancies ($Vac_A$ and $Vac_{A'}$, yellow and green bars), anion vacancies ($Vac_B$), and cation anti-site clusters ($A_{A'}+A'_A$). The ternaries considered exhibit a composition of AA'$_2$B$_2$. Hashed regions signify variation in vacancy formation energies with change in the chemical potential of the species removed.

## 4 Discussion

Using DFT calculations, we have explored the binary, ternary, and quaternary divalent-metal-based pnictide chemical space as candidate beyond-Si photovoltaic materials in this work. Specifically, we considered a set a twenty compounds, comprising A$_3$B$_2$ binaries, AA'$_2$B$_2$ ternaries, and AA'A"B$_2$ quaternaries (A, A', A" = Ca, Sr, or Zn; B = N or P) and quantified their ground state configurations, band gaps, 0 K thermodynamic stability, and formation energies of select point defects. Importantly, we identified SrZn$_2$N$_2$, SrZn$_2$P$_2$, and CaZn$_2$P$_2$ as candidate materials based on the properties calculated, besides ZnCa$_2$N$_2$ and ZnSr$_2$N$_2$, which are susceptible to Vac$_N$ formation.

For the pnictides whose structures were unavailable in the ICSD, we calculated the ground state configuration by considering possible structures using other binary and ternary pnictide unit cells that were available as templates. However, we did not consider the possibility of cation arrangements forming superlattices over larger distances, which requires the consideration of larger supercells (instead of smaller unit cells), thus significantly increasing computational costs. Thus, for the candidate ternaries that we have identified (namely, SrZn$_2$N$_2$, SrZn$_2$P$_2$, and CaZn$_2$P$_2$), it may be useful to explore the possibility of superlattice formation. Also, we did not consider the anti-bixbyite structure of Ca$_3$N$_2$/Zn$_3$N$_2$ as templates for the ternary and quaternary structures since the structure has 48 symmetrically distinct cation sites and decorating, say, 16 sites each of Ca, Sr, and Zn is computationally intractable. Hence, we do not rule out our candidate ternaries from crystallizing in anti-bixbyite-based structures.

We carried out DOS calculations using SCAN and HSE06 because the former typically underestimates $E_g$. Since hybrid functionals include a portion of the exact exchange, their band gap estimates are typically better due to the consequent reduction in self-interaction errors compared to semi-local functionals. Indeed, previous studies have reported better agreement in band gap estimates with HSE06 vs. experiments compared to SCAN in kesterite photovoltaics.[46,27] However, hybrid functionals do not guarantee the best theoretical estimates of $E_g$, which typically require quasi-particle calculations, such as the single-shot GW calculation (i.e., G$_0$W$_0$), which has shown excellent agreement with experimental photoemission spectroscopy/inverse photoemission spectroscopy measurements before.[49,50] However, the computational costs of GW calculations are even higher than hybrid functionals, which is the reason we did not pursue such calculations for all the twenty compounds considered in our work.

The synthesis of nitrides is difficult given their tendency to decompose at higher temperatures due to lower formation energies (compared to oxides),[14] the strong triple bond of N$_2$ molecule, and stringent conditions of oxygen and water-free conditions to achieve high purity.[51,52] Hence, we have focused solely on compounds that are stable at 0 K since we expect the synthesis to be easier compared to metastable compounds ($E^{hull} \leq 30$ meV/atom), which may require higher temperatures and/or pressures. So far, synthesis of Ca and Zn based ternary nitrides have been performed via solid state techniques, using binary Ca$_3$N$_2$ and Zn$_3$N$_2$ as precursors,[14] where ZnCa$_2$N$_2$ required Ar-atmosphere and CaZn$_2$N$_2$ needed high pressure and an elevated temperature.[14] Besides synthesis difficulties, the air and moisture sensitivity of the candidate compounds (both nitrides and phosphides), particularly during operation as a photovoltaic, and the consequent impact on solar absorption efficiencies remains to be quantified. In the case of phosphides, the possible formation of the highly-toxic phosphine gas, either during synthesis or during operation, is also a concern.[53,54] Another challenge with pnictides will be the ability to form *p-n* homojunctions, instead of relying on other materials to form heterojunction-based devices.[42]

While DFT often does not provide quantitative accuracy for defect formation energies, previous studies have shown that SCAN can provide an upper bound (compared to other semi-local functionals) on defect formation energies, with qualitative trends being similar across different functionals for a given compound.[27,31] For reducing quantitative error in defect formation energies, we need further improvements in the XC functional and likely shoulder higher computational costs. Nevertheless, we expect all candidate pnictides shown in **Figure 4** to be significantly more resistant to point defect formation, such as cation vacancies, cation anti-sites, and anion vacancies, compared to other compound semiconductors, such as $Cu_2ZnSnS_4$.

## 5 Conclusion

The design of new semiconducting beyond-Si materials that can achieve reasonable solar-to-electricity conversion efficiencies, and that consist of relatively abundant and non-toxic elements is a key ingredient in moving towards non-fossil-fuel-based sources of energy. In this work, we have used DFT calculations to systematically explore the pnictide chemical space, including binary $A_3B_2$, ternary $AA'_2B_2$, and quaternary $AA'A''B_2$ compounds, where A, A', and A'' are non-redox-active divalent cations (namely Ca, Sr, or Zn), and B is either N or P. In total, we considered a set of 20 pnictide compositions, out of which 10 were experimentally known structures and we created the initial (theoretical) structures of the remaining 10 compounds using experimentally-available template structures. Upon identifying the DFT-calculated ground state configurations of all 20 compositions, we evaluated the band gaps, 0 K thermodynamic stability, and the resistance to formation of cation/anion vacancies and cation anti-sites. For treating the electronic exchange and correlation, we used a combination of the SCAN and HSE06 functionals, with SCAN used for structure relaxations and HSE06 for band gap evaluations. Wherever possible, we benchmarked our calculated values with those available in the literature. Importantly, we found a set of five pnictide compositions that exhibit HSE06-calculated band gaps in the optimal (1.1-1.5 eV) range, are thermodynamically stable at 0 K, and are resistant to formation of point defects (formation energy > 0.4 eV), namely, $SrZn_2N_2$, $SrZn_2P_2$, $CaZn_2P_2$, $ZnCa_2N_2$, and $ZnSr_2N_2$. Among the candidates identified, we found SrZn$_2$N$_2$, SrZn$_2$P$_2$, and CaZn$_2$P$_2$ to be particularly resistant to forming point defects (formation energies >1 eV), while the other compounds (ZnCa$_2$N$_2$ and ZnSr$_2$N$_2$) are susceptible to Vac$_N$ formation (0.42-0.47 eV). We hope that our study will ignite further interest in the pnictide chemical space and enable the practical realization of some of the candidates identified in this work as beyond-Si photovoltaics.

**Electronic Supporting Information**

PAW potentials used, SCAN and HSE06 calculated DOS data, range of chemical potentials used for defect calculations, and structural information on all theoretical structures considered.

**Data Availability**

All the computational data presented in this study are freely available to all on our GitHub repository (https://github.com/sai-mat-group/pv-nitrides).

**Acknowledgments**

G.S.G acknowledges financial support from the Indian Institute of Science (IISc) Seed Grant, SG/MHRD/20/0020 and SR/MHRD/20/0013. J.K. thanks the Ministry of Human Resource Development, Government of India, for financial assistance. The authors acknowledge the computational resources provided by the Supercomputer Education and Research Centre, IISc, for enabling some of the density functional theory calculations showcased in this work.